# Non-Magnetic Cloak without Reflection

Wenshan Cai[1], Uday K. Chettiar[1], Alexander V. Kildishev[1],
Graeme W. Milton[2], and Vladimir M. Shalaev[1,*]

[1]*School of Electrical and Computer Engineering and Birck Nanotechnology Center,*
*Purdue University, West Lafayette, Indiana 47907, USA*

[2]*Department of Mathematics, University of Utah, Salt Lake City, Utah 84112, USA*

[*] E-mail: shalaev@purdue.edu

**Abstract:**
In an electromagnetic cloak based on a transformation approach, reduced sets of material properties are generally favored due to their easier implementation in reality, although a seemingly inevitable drawback of undesired reflection exists in such cloaks. Here we suggest using high-order transformations to create smooth moduli at the outer boundary of the cloak, therefore completely eliminating the detrimental scattering within the limit of geometric optics. We apply this scheme to a non-magnetic cylindrical cloak and demonstrate that the scattered field is reduced substantially in a cloak with optimal quadratic transformation as compared to its linear counterpart.

Recently, increasing attention has been focused on creating an electromagnetic cloak of invisibility based on various schemes, including anomalous localized resonance,[1,2,3] dipolar scattering cancellation,[4,5] tunneling light transmittance,[6] sensors and active sources,[7] and coordinate transformation.[8,9] The transformation approach, which generalized a similar idea on cloaking of conductivity,[10,11] has triggered enormous interest because the proposed device is supposed to render a macroscopic object invisible, and the design is not sensitive to the object that is being masked. Further theoretical treatments of the problem were recently presented.[12,13,14] The proof-of-concept demonstration of such a cloak operating at microwave frequency was experimentally reported,[15] which showed that the detour of waves and the reduced scattering from the cloaked object were indeed possible.

As pointed out in our recent paper,[16] the design for the microwave cloak[15] is not directly adaptable to optical frequencies due to the intrinsic limits of the scaling of split-ring-resonators (SRRs) that would be necessary to exhibit a magnetic response in the optical range.[17] Replacing the SRRs with other high-frequency magnetic elements like coupled nanorods,[18] nanoplates[19] or nanostrips[20,21] is also questionable due to the lossy nature of such plasmonic magnetic structures. To realize an optical cloak that is inherently compatible with the natural inertness of magnetic responses in the visible range, we proposed the physical principle and implementation recipe of a non-magnetic optical cloak as detailed in Ref. 16. Another version of a non-magnetic cloak was suggested in Ref. 9, where an isotropic medium extending to infinity with a designed gradient in the refractive index was used to construct a two-dimensional cloak, which is exact in the geometric optics limit.

In our proposed non-magnetic optical cloak,[16] a cylindrical region $r' \leq b$ is compressed into a concentric cylindrical shell $a \leq r \leq b$, and anisotropic material with a set of reduced parameters without any magnetism requirement is utilized to form the shell. The idea of using reduced parameters to achieve an approximate cloak was originated in Ref. 22. The detrimental effect of using the non-magnetic reduced set as compared to the perfect set is the non-zero reflection due to impedance mismatch at the outer surface of the cloaking system. For this non-magnetic cloak, the impedance $Z = \sqrt{\mu_z/\varepsilon_\theta}$ is distinct from unity, being $Z = 1 - a/b$ at the outer boundary $r = b$.

In this Letter, we use a high-order coordinate transformation to eliminate the undesired scattering from the outer boundary of the cylindrical cloaking system while sustaining the important non-magnetic feature for optical frequencies. Mathematically, there are countless ways to convert a two-dimensional cylindrical or three-dimensional spherical region $r' \leq b$ into a concentric shell occupying the space $a \leq r \leq b$. Up to now in all the reported literature on this topic,[8,15,16,22,23] a linear transformation function is used for this purpose:

$$r = \left(1 - a/b\right)r' + a \qquad (1)$$

while the other two coordinates ($\theta$, $z$ for a cylindrical system and $\theta$, $\phi$ for a spherical system) are kept untouched. This linear transformation, although straightforward and intuitive, prohibits any flexible control of the associated moduli. As a result, impedance mismatching and undesired reflection are inevitable for cloaks using any form of reduced parameters, including the demonstrated microwave cloak in Ref. 15 for TE incidence and the designed optical cloak in Ref. 16 for TM polarization.

This scenario, however, can be dramatically changed when using a high-order transformation instead of the linear one in Eq. (1). The basic idea is outlined as follows. We may conceive any possible transformation function $r = g(r')$ from $(r', \theta', z')$ to $(r, \theta, z)$ in order to compress the region $r' \leq b$ into a concentric shell of $a \leq r \leq b$ in a cylindrical (or spherical) coordinate system. We allow several flexible variables in the expression of $g(r')$ for further adjustments as detailed later. The transformation function $g(r')$ must fulfill boundary confinements such as $g(0) = a$ and $g(b) = b$. Moreover, monotonicity of the transformation is required such that $\partial g(r')/\partial r' > 0$ when $0 \leq r' \leq b$. With the given form of the function $g(r')$, we calculate the Jacobian matrix for this coordinate change, and the permittivity $\varepsilon(r, \theta, z)$ and permeability $\mu(r, \theta, z)$ tensors can be determined as well based on the techniques described in Refs. 8, 23. After that, the corresponding reduced parameters are derived following the approach discussed in Refs. 15, 16. With all the anisotropic material properties available, the impedance at the outer boundary of the cloak $Z|_{r=b}$ is expressed using the geometrical parameters and the flexible variables in the transformation function $g(r')$. By setting $Z|_{r=b}$ equal to unity, we can fix the function $g(r')$ together with all the material properties, and the reflectionless cloak with reduced parameters is achieved.

Following the implementation guidance described above, we propose a non-magnetic cloak with zero reflection in a cylindrical system based on a high-order transformation approach. As a starting point, we consider a second order (quadratic) transformation function in the following form with one flexible parameter $p$:

$$r = g(r') = \left[1 - a/b + p(r' - b)\right]r' + a \tag{2}$$

and monotonicity requires $|p| < (b-a)/b^2$.

First we calculate the transformation coefficients $g_{ij} = \sum_l (\partial x_l/\partial q_i)(\partial x_l/\partial q_j)$ from a Cartesian mesh ( $x = r'\cos\theta'$ ; $y = r'\sin\theta'$ ; $z = z'$ ) to the transformed system ( $r = g(r'); \theta = \theta'; z = z'$ ). The diagonal elements of the transformation matrix are:

$$h_r = \sqrt{g_{rr}} = \left[\left(\frac{\partial x}{\partial r'}\Big/\frac{\partial g(r')}{\partial r'}\right)^2 + \left(\frac{\partial y}{\partial r'}\Big/\frac{\partial g(r')}{\partial r'}\right)^2\right]^{1/2} = \frac{1}{p(2r'-b) + 1 - a/b} \tag{3a}$$

$$h_\theta = \sqrt{g_{\theta\theta}} = \left[\left(\frac{\partial x}{r\partial\theta}\right)^2 + \left(\frac{\partial y}{r\partial\theta}\right)^2\right]^{1/2} = \frac{r'}{r} \tag{3b}$$

$$h_z = \sqrt{g_{zz}} = 1 \tag{3c}$$

Based on these elements in the transformation matrix, we obtain the following anisotropic permittivity and permeability components for the construction of an ideal electromagnetic cloak ($a < r < b$):

$$\varepsilon_r = \mu_r = \frac{h_\theta h_z}{h_r} = \frac{r'}{r}\left[p(2r'-b) + 1 - a/b\right] \tag{4a}$$

$$\varepsilon_\theta = \mu_\theta = \frac{h_r h_z}{h_\theta} = \frac{1}{\varepsilon_r} \tag{4b}$$

$$\varepsilon_z = \mu_z = \frac{h_r h_\theta}{h_z} = \frac{r'}{r}\frac{1}{p(2r'-b) + 1 - a/b} \tag{4c}$$

For the linear transformation with $p = 0$, the transformation and material parameters in Eqs. (2) and (4) reduce to the simple forms given in Refs. 15, 22. Note that for a closed-form expression of the parameters in ($r, \theta, z$) space, all $r'$ in the formulae above should be replaced by

$$r' = \frac{1}{2p}\left[\sqrt{(1-a/b-pb)^2 + 4p(r-a)} - (1-a/b-pb)\right] \tag{5}$$

From the expressions in Eq. (4) we can see that the impedance at the outer boundary of the cloak $r = b$ is perfectly matched, that is, $\sqrt{\mu_\theta/\varepsilon_z}\Big|_{r=b} = \sqrt{\mu_z/\varepsilon_\theta}\Big|_{r=b} = 1$, which is an important feature of an ideal cloak.

As stated in Ref. 16, a non-magnetic cloak is of particular interest at optical frequencies due to the absence of optical magnetism in nature. We focus on TM incidence with the magnetic field polarized along the $z$ axis. In this case only $\mu_z$, $\varepsilon_r$ and $\varepsilon_\theta$ enter into Maxwell's equations. To simplify the $\varepsilon$ and $\mu$ tensors while maintaining the dispersion relationship and the wave trajectory inside the cylindrical shell, in Eq. (4) we multiply $\varepsilon_r$ and $\varepsilon_\theta$ by the value of $\mu_z$ and obtain the following reduced set of non-magnetic cloak parameters:

$$\varepsilon_r = \left(r'/r\right)^2;\ \varepsilon_\theta = \left[p(2r'-b)+1-a/b\right]^{-2};\ \mu_z = 1. \tag{6}$$

Again, for a closed-form solution, all $r'$ above should be replaced by the expression in Eq. (5). Note that for all valid quadratic transformations, $\varepsilon_r$ varies from 0 at the inner surface to 1 at the outer side. The azimuthal permittivity $\varepsilon_\theta$ is constant only in the linear transformation case with $p = 0$.

In this new non-magnetic cloak with a quadratic transformation, the impedance at the outer boundary $r = b$ is given as:

$$Z\big|_{r=b} = \sqrt{\mu_z/\varepsilon_\theta}\Big|_{r=b} = p(2r'-b)+1-a/b \tag{7}$$

For a linear transformation with $p = 0$, the impedance above reduces to $Z = 1 - a/b$ as shown in an earlier part of this Letter. Since $Z\big|_{r=b}$ is a strong function of the transformation variable $p$, we can choose the appropriate $p$ to match $Z\big|_{r=b}$ to the surrounding impedance. By setting $Z\big|_{r=b} = 1$, we obtain the following optimal transformation:

$$p = a/b^2 \tag{8a}$$

$$r = g(r') = \left[(a/b)(r'/b - 2) + 1\right]r' + a \tag{8b}$$

and all non-magnetic material properties can be determined consequently using Eqs. (5) and (6). To make sure that the transformation is monotonic, we require $|p| < (b-a)/b^2$. Hence, with quadratic transformations, the non-magnetic cylindrical cloak without reflection is only possible for a relatively thick cloak with a shape factor $a/b < 0.5$.

The optimal quadratic transformation also indicates a smooth modulus at the outer boundary. The transformation function in Eq. (2) results in

$$\frac{\partial r}{\partial r'} = \frac{\partial g(r')}{\partial r'} = p(2r' - b) + 1 - a/b \tag{9}$$

Therefore $p = a/b^2$ implies $\left(\partial r/\partial r'\right)\big|_{r=b} = 1$, which removes the discontinuity at the outer surface $r = b$ after the high-order transformation.

As an example, Fig. 1 shows the anisotropic material properties of two non-magnetic cylindrical cloaks with $p = a/b^2$ (optimal quadratic transformation) and $p = 0$ (linear transformation) respectively. The shape factor in this example is $a/b = 0.31$. In the optimal quadratic case, all three material parameters $\varepsilon_r$, $\varepsilon_\theta$ and $\mu_z$ are equal to unity at the outer boundary $r = b$, which perfectly matches the surrounding vacuum parameters.

To compare the performance of the non-magnetic cloaks with different transformations methods, we conduct field-mapping simulations using the finite-element package COMSOL Multiphysics. The cloaking systems are operated at $\lambda$ = 632.8 nm with the same shape factor as that in Fig. 1. Both the size of the cloaks ($b$ = 2 μm) and the simulation domains are substantially larger than the operational wavelength. The object hidden inside the cloaks is an ideal metallic cylinder with a radius the same as that of the inner surface. In all simulations, a plane wave propagates from the left-hand side of the simulation domain with its electric field polarized normal to the axis of cylindrical system. In Fig. 2, under four circumstances we plot the magnitudes of the normalized scattered field, which is obtained by subtracting the incident field from the total field outside the cloak devices and then normalizing it to the incident field itself. Fig. 2a shows the scattering property of the object (metal cylinder) itself without any cloaking system. The strong scattering to the right-hand side of the system corresponds to the obvious shadow cast behind the object. The scattered field outside an ideal cloak is illustrated in Fig. 2b, which is essentially zero in magnitude. The results of non-magnetic cloaks for both the linear transformation with $p = 0$ and the optimal quadratic case with $p = a/b^2$ are illustrated in Fig. 2c and 2d, respectively. The linear case exhibits an evident scattering pattern from the outer boundary of the system because of the impedance mismatch. On the other hand, the quadratic transformation function results in negligible reflectance from the cloaking system. The minor scattering around the quadratic cloak results from the gradient in the impedance inside the cylindrical shell. The figure of merit for cloaking (defined as the ratio of the scattering cross-sections without and with the cloak) is about 10 for the considered quadratic cloak, and it increases towards infinity with the size of the cloaking system. The cloaking performance of an optimal quadratic non-magnetic cloak, like that of the cloak of Leonhardt,[9] is exact as an ideal cloak within the realm of geometric optics.

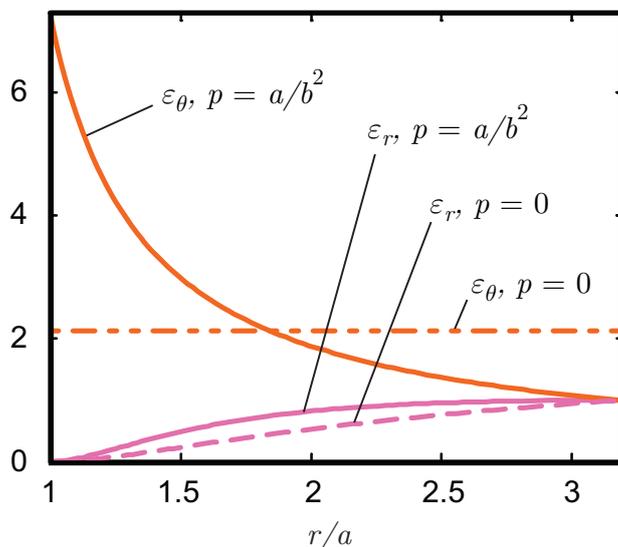

Fig. 1. Anisotropic material parameters $\varepsilon_r$, $\varepsilon_\theta$ of two non-magnetic cloaks with $p = a/b^2$ (optimal quadratic transformation, solid lines) and $p = 0$ (linear transformation, dashed lines). $\mu_z$ equals unity in both cases. The shape factor ($a/b$) in this example is 0.31.

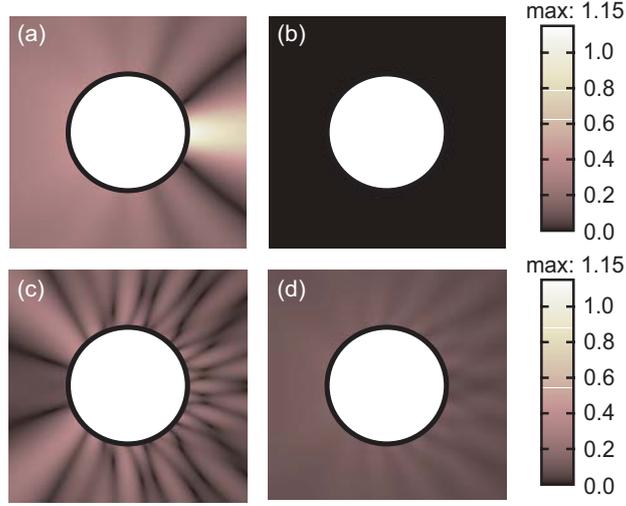

Fig. 2. Full-wave field-mapping simulations of the magnitudes of normalized scattered field for a metal cylinder inside (a) a vacuum without any cloak; (b) an ideal linear cloak; (c) the linear non-magnetic cloak with $p = 0$ and (d) the optimal quadratic cloak with $p = a/b^2$.

    To clearly illustrate the scattering and directivity properties of different cloaking systems, in Fig. 3 we plot the scattering radiation patterns corresponding to the four cases in Fig. 2. The curves in Fig. 3 show the energy flow in the radial direction normalized by the maximum value in the non-cloaked case at a boundary outside the outer surface of the cloaks. The shadow and reflection from the object without any cloak are obvious as indicated by the strong lobes in both the forward and backward directions. In the ideal cloaking system, the scattering energy flow is zero, which is indicated by the solid inner circle in Fig. 3. The linear cloak with reduced parameters gives rise to a noticeable and strongly directional scattering pattern due to the impedance difference at the boundary. In the non-magnetic quadratic cloak, the overall scattering is much less significant. The peak value of the radial Poynting vector in the quadratic cloak is more than six times smaller than that of its linear counterpart. Moreover, the directivity in the scattering pattern is substantially suppressed, which is an important feature of a quasi-ideal cloak.

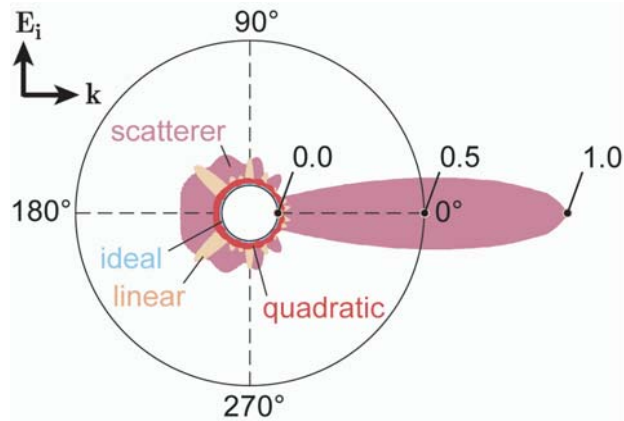

Fig. 3. The scattering patterns from the four cases corresponding to those in Fig. 2. Purple: a metal cylinder (scatterer) with no cloak; blue: the cylinder with the ideal cloak; orange: the cylinder with a linear non-magnetic cloak; red: the cylinder with an optimal quadratic non-magnetic cloak.

In conclusion, we proposed an electromagnetic cloak using high-order transformation functions to create smooth rather than discontinuous moduli at the outer boundary of the cloaking shell. By this approach, the undesired reflection from the system is completely eliminated within the limit of geometric optics, even for cloaks using non-magnetic materials to simplify the implementation. We applied this scheme to the non-magnetic cylindrical cloak and demonstrated that the scattered field from the outer boundary is reduced by almost an order of magnitude in a cloak with optimal quadratic transformation comparing to that with the usual linear compression. This non-magnetic cloak with zero reflection brings us one step closer to realizing an ideal and exact cloaking device at optical frequencies.

**Acknowledgement**

This work was supported in part by ARO-MURI award 50342-PH-MUR and by National Science Foundation through grant DMS-0411035.